# Coupling of Electronic Transitions to Ferroelectric Order in a 2D Semiconductor

Chun-Ying Huang,[1,*] Daniel G. Chica,[1,*] Zhi-Hao Cui,[1,*] Taketo Handa,[1] Morgan Thinel,[1,2] Nicholas Olsen,[1] Yufeng Liu,[1] Michael E. Ziebel,[1] Guiying He,[3,4] Yinming Shao,[2,6] Connor A. Occhialini,[2,5] Jonathan Pelliciari,[5] Dmitri N. Basov,[2] Matthew Sfeir,[3,4] Abhay Pasupathy,[2] Valentina Bisogni,[5] David R. Reichman,[1,†] Xavier Roy,[1,†] and Xiaoyang Zhu.[1,†]

[1]Department of Chemistry, Columbia University, New York, NY 10027, USA
[2]Department of Physics, Columbia University, New York, NY 10027, USA
[3]Department of Physics, Graduate Center, City University of New York, New York, NY 10016, USA
[4]Photonics Initiative, Advanced Science Research Center, City University of New York, New York, NY 10031, USA
[5]National Synchrotron Light Source II, Brookhaven National Laboratory, Upton, NY 11973, USA
[6]Department of Physics, Pennsylvania State University, University Park, PA, 16802, USA

**Abstract.** A ferroelectric material often exhibits a soft transvers optical (TO) phonon mode which governs it phase transition. Charge coupling to this ferroelectric soft mode may further mediate emergent physical properties, including superconductivity and defect tolerance. However, direct experimental evidence for such coupling is scarce. Here we show that a photo-launched coherent phonon couples strongly to electronic transitions across the bandgap in the van der Waals (vdW) two-dimensional (2D) ferroelectric semiconductor $NbOI_2$. Using terahertz time-domain spectroscopy and first-principles calculations, we identify this mode as the TO phonon responsible for ferroelectric order. This exclusive coupling occurs only with above-gap electronic transition and is absent in the valence band as revealed by resonant inelastic X-ray scattering. Our findings suggest a new role of the soft TO phonon mode in electronic and optical properties of ferroelectric semiconductors.

---

[*] These authors contributed equally.
[†] To whom correspondences should be addressed. DRR: drr2103@columbia.edu; XR: xr2114@columbia.edu; XYZ: xyzhu@columbia.edu

**Introduction**

In conventional polar semiconductors where lattice displacements are described by harmonic oscillators, the coupling between charge carriers and nuclei is dominated by longitudinal optical (LO) phonons. This is the so-called Fröhlich interaction which describes the long-range interaction between a charge carrier and the effective dipoles along the LO phonon coordinate[1,2]. By contrast, the Coulomb interactions of transverse optical (TO) phonon mode have a negligible contribution to the long wavelength electric field due to the lack of macroscopic polarization. In ferroelectric materials, however, a unique potential energy surface associated with the ferroelectric TO phonon mode appears and is characterized by a double well with minima at two non-zero electric polarizations of equal magnitudes but opposite signs[3,4]. Such a TO mode is typically labeled as "soft" as the double well flattens, and the corresponding frequency decreases with increasing temperature towards the phase transition. The strong interaction between charge carriers and the soft TO mode is believed to be responsible for charge pairing in a class of ferroelectric superconductors[5,6] and for the efficient screening of charge carriers in para- and ferroelectric semiconductors in the ferroelectric polaron model [7–11]. Charge carrier scattering with polar nano domains resulting from fluctuations in the soft TO mode is believed to determine charge carrier mobility in ferroelectric dielectrics [12–16].

Despite interests in carrier-TO phonon coupling in ferro- and paraelectric materials, our current understanding is primarily based on conceptual models, computations, and indirect experimental evidences [7–16]. Here we investigate coupling of phonons to electronic transitions in the 2D vdW ferroelectric semiconductor $NbOI_2$, which exhibits in-plane ferroelectric polarization and an optical bandgap in the visible range[17,18]. The vdW nature of $NbOI_2$ leads to little interlayer interaction[19], thus allowing us to probe the interactions primarily in 2D, in which coulomb interaction is strong. Using pump-probe coherent phonon spectroscopy, we find that electronic transition across the bandgap couples exclusively to a phonon mode at 3.1 THz, which is identified as the TO phonon in terahertz time-domain spectroscopy (THz-TDS). Theoretical modeling based on density functional theory reveals that this TO phonon mode corresponds to stretching of the Nb−O polar bonds responsible for ferroelectric order. Using resonant inelastic X-ray scattering (RIXS) spectroscopy[20–22], we further establish that the ferroelectric TO mode couples exclusively with electronic transition above the band gap, a mechanism distinct from valence-band electron-phonon coupling.

**Results and Discussions**

We grow millimeter-size single crystal NbOI$_2$ with optically flat surfaces (Supporting Information, Figure. S1a), using the chemical vapor transport method[17] and confirm the phase purity through powder X-ray diffraction (Figure S1b). NbOI$_2$ adopts a vdW layered monoclinic crystal structure in the $C_2$ space group (Figure 1a). Along the *b*-axis, the Nb atoms are displaced from the center position, resulting in unequal Nb−O bond lengths and spontaneous polarization along the *b*-axis. Scanning tunneling microscopy image (Figure. 1b) of the crystal surface freshly cleaved in vacuum confirms the rectangular in-plane structure with lattice parameters consistent with those obtained by single-crystal X-ray diffraction (Figure S2). The anisotropic in-plane crystal structure gives rise to an anisotropic electronic structure, as revealed by differential reflectance (Δ*R*/*R*) spectroscopy (Figure 1c). The Δ*R*/*R* spectrum along the *c*-axis reveals across-gap transitions at 2.3 eV and 2.6 eV[17]; these transitions are absent along the *b*-axis. A weak onset in absorption is visible at ~1.1 eV (Figure. S3), likely due to an indirect optical transition. By performing high-temperature second harmonic generation (SHG) and Raman spectroscopy, we also find the ferroelectric phase remains robust up to 573 K (Figure S4).

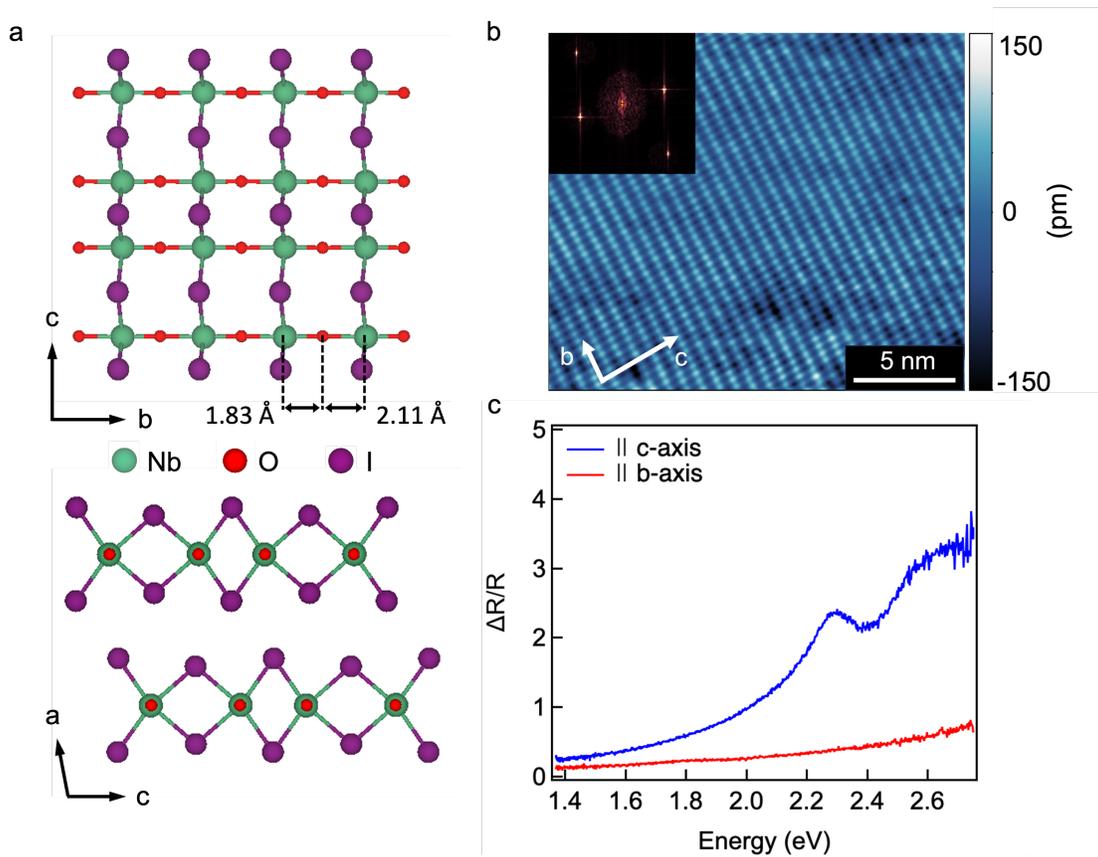

**Figure 1. a.** Crystal structure of NbOI$_2$ viewed along *a*-axis (upper panel) and *b*-axis (lower panel). **b.** Scanning tunneling microscopy image of cleaved NbOI$_2$ single crystal at room temperature. Inset shows the FFT of the topography. **c.** Differential reflectance spectra of 7-layers NbOI$_2$ on a *z*-cut quartz with incident light polarized along either crystallographic *c*- or *b*-axis at room temperature. Δ*R* is defined as $R_{sample} - R_{substrate}$ in steady state and $R$ in the dominator is $R_{substrate}$.

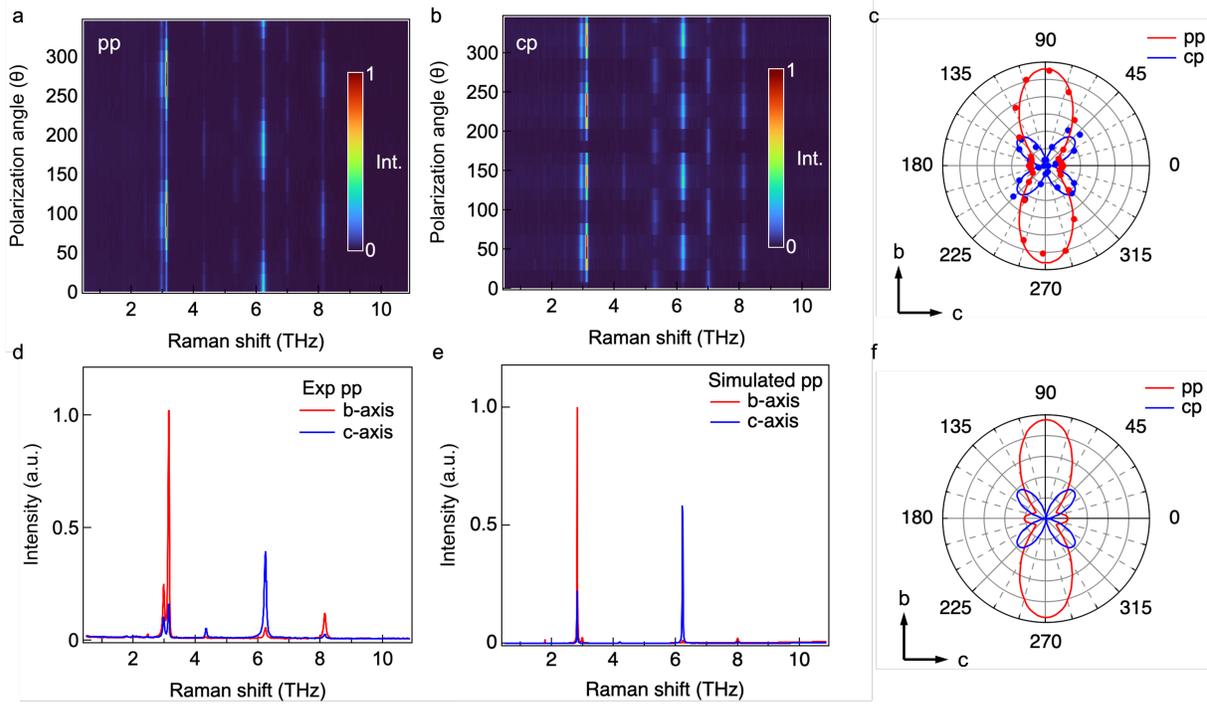

**Figure. 2.** (**a**, **b**) Polarization-angle resolved Raman spectra of NbOI$_2$ at RT with incident polarization in the *yz* plane (*y* = *b*-axis, *z* = *c*-axis) and scattering polarization parallel (**a**) or perpendicular (**b**) to incident polarization. 0 degree is referenced to crystallographic *c*-axis. Color bar represents the Raman scattering intensity (normalized to the highest intensity for clarity). pp represents parallel polarization and cp represents cross polarization. **c**. Polar plot corresponding to spectral cut from (**a**) and (**b**) at 3.137 THz mode (P2 mode). Solid lines are fitting curves obtained by the global fit to both pp and cp data with equation 1 and 2 in Supplementary Text, respectively. **d**. Parallel polarized Raman spectrum with incident polarization aligned to either the *b* or *c*-axis. **e.** Simulated Raman spectra at 300 K with parallel polarization. **f.** Simulated polar plot of the phonon mode at a calculated frequency of 2.830 THz. An empirical term, $\cos\varphi_{c-b}$, that was deduced from the tensor fit of the experimental P2 mode data and describing the phase difference between two tensor elements were applied to the simulation to count the finite absorption of incident light.

To determine the phonon modes in NbOI$_2$, we performed polarization-resolved Raman spectroscopy with incident polarization in the *bc* (*yz*) plane and scattered polarization parallel (pp) or perpendicular (cp) to incident polarization. The polarization angle (*θ*) is referenced to the crystallographic *c*-axis. Figure 2a,b present pseudo-color plots of Raman spectra with pp and cp polarizations, respectively, along with polar plots of the Raman peak intensities for the strongest mode at 3.137 THz (104.5 cm$^{-1}$, labeled as P2 in main text) in Figure 2c. This mode is highly anisotropic and preferentially allowed along the *b*-axis for pp (red). For cross polarization (blue), the polar plot is symmetric along the diagonals between the *b*- and *c*-axes. We conduct Raman tensor analysis to assign symmetries to the Raman-active phonons (see Supporting Information for details)[23–26]. We extract line-cuts of the spectra at $\theta$ = 0º (*c*-axis) and 90º (*b*-axis) to show the anisotropy of the Raman-active modes (Figure 2d). A comparison of the experimental data with theoretical Raman spectra calculated from first-principles density functional theory (DFT) in Figure 2e shows good agreement across the measured range (see Methods and Table S1 for full assignment of calculated Raman modes). In particular, an excellent agreement can be confirmed for the anisotropy of the main phonon mode (this mode is 2.83 THz in computation, corresponding the P2 mode in experiments), as shown in Figure 2f. The discrepancy in the position of the small shoulder peak—down-shifted in experiment and up-shifted in simulation—is attributed to anharmonicity, which can be approximately corrected by other methods, for example, in ref[27]. As detailed below, the main phonon mode, P2, corresponds to the ferroelectric TO phonon and strongly couples to photogenerated carriers.

We carry out ultrafast pump-probe coherent phonon spectroscopy (CPS) to probe how electronic transitions couple to phonons in NbOI$_2$. We use a ~150 fs excitation laser pulse centered at $h\nu_1$ = 2.58 eV, which is resonant with the second above-gap transition, with light incidence normal to the surface and polarization parallel to the *c*-axis. Figure 3a and 3b show the transient reflectivity spectrum ($\Delta R/R_0$, where $R_0$ is reflectance without pump and $\Delta R$ is the pump-induced change in reflectance), probed by a broadband pulse ($h\nu_2$ ~ 1.46−2.50 eV) polarized parallel to the *c*- or *b*-axis, respectively. In transient reflectance probed along the optical-transition allowed direction (*c*-axis), there is a clear signature of exciton bleaching at ~2.20−2.35 eV, with periodic modulation of the signal arising from coherent phonon(s). The clear oscillatory signals at delay times > 30 ps without subtracting the electronic signal demonstrates the long coherence time at room temperature with a decay constant of ~30 ps (also see Figure S6 for short-time Fourier transform analysis). The full kinetic trace of exciton bleaching (Figure S7) can be described by a long decay time constant of $\tau_d$ = 5.6 ± 1.2 ns. After removal of the incoherent decay component, the Fourier transform (FT) of the residue coherent signals (Figure 3c and 3d) give a single frequency centered at 3.125 ± 0.001 THz with a full-width-at-half-maximum (FWHM) of 0.038 ± 0.001 THz (Figure S7a). This oscillatory frequency closely matches the frequency of the P2 phonon mode observed in polarization-angle resolved Raman spectra.

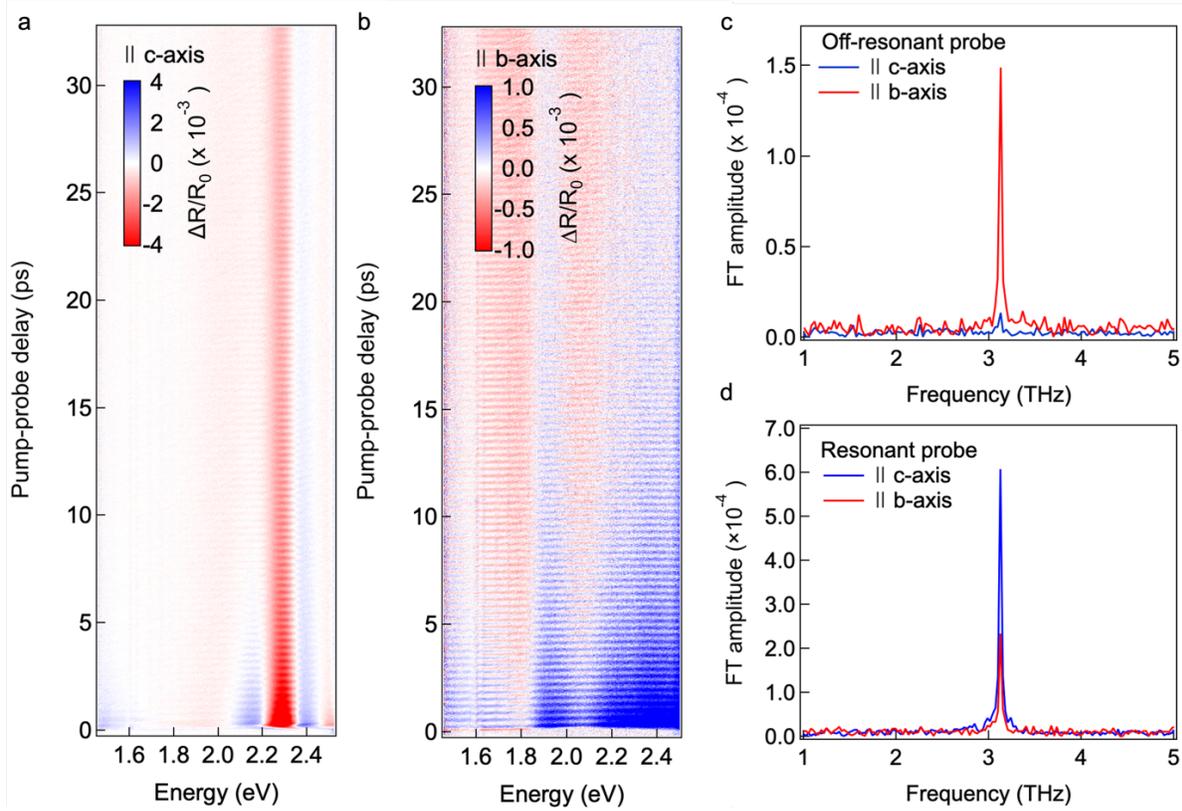

**Figure 3.** (**a,b**). Transient reflectivity spectra of NbOI$_2$ measured at RT for probe polarization parallel to (**a**) *c*- or (**b**) *b*-axis. Color bar indicates the magnitude of $\Delta R/R_0$. **c.** FT of the oscillatory components extracted from (**a,b**) at off-resonant probe region (1.55 eV). **d.** FT of the oscillator components extract from (**a,b**) at resonant probe region (2.30 eV).

The polarization dependencies of the CPS spectra for resonant and off-resonant probes show key differences. When probed with a photon energy below the excitonic transition (off-resonant, Figure 3c), the phonon mode dependence is highly anisotropic. The FT amplitude of the 3.125 THz peak is prominent with probe polarization along the *b*-axis ($I_b$) and more than one order of magnitude lower with polarization along the *c*-axis ($I_c$). The ratio $I_c/I_b \sim 0.1$ from CPS data (Figure 3c) is close to the ratio $I_c/I_b = 0.16$ obtained from Raman spectroscopy (Figure 2c), consistent with the symmetry of the Raman tensor. Moreover, the FWHMs of the peaks measured by CPS (0.038 ± 0.001 THz),) and Raman spectroscopy (0.039 ± 0.002 THz) are essentially the same. We conclude that with a non-resonant, below-gap probe, the P2 phonon mode revealed in CPS is determined by the same Raman tensor[28,29] as in conventional Raman spectroscopy.

By contrast, when we use a resonant probe (Figure 3d), the CPS spectrum reveals the breaking of the selection rule dictated by the Raman tensor. The weak coherent phonon amplitude for probe polarization along the *c*-axis becomes dominant with the resonant probe. The FT amplitude ratio $I_c/I_b$ for the P2 mode increases from ~0.1 with the off-resonant probe to 2.6 with the resonant probe. We attribute the emergence of the intense symmetry-forbidden P2 mode to the strong coupling between the P2 mode and the above-gap electronic transition. This coupling can result from the displacement along the P2 phonon coordinate in the potential energy surface corresponding to the optical transition due to the strong electron-phonon coupling associated with the conduction band. Since above-gap electronic transition is related to carrier generation, this strong coupling suggests polaronic feature of the photoexcited carriers[30,31]. As we show below, the P2 mode is identified as a TO phonon mode, the polaronic nature of the photo-generated carriers are beyond the Fröhlich model.

To reveal the nature of the P2 phonon mode, we performed terahertz time-domain spectroscopy (THz-TDS) measurements in transmission geometry on a free-standing $NbOI_2$ crystal with thickness ~2.6 μm (Figure S8). Because the transverse nature of electromagnetic wave only allows excitation of TO phonon modes, THz-TDS directly probes TO phonons. Figure 4a and 4b show the time-domain waveforms and their corresponding FT spectra along the *b*- (red) and *c*- (blue) crystallographic axes, respectively. The strong THz absorption peak is observed at 3.130 ± 0.008 THz (Figure 4b), and this frequency is the same as that of the exclusive mode observed in CPS (reproduced in Figure 4c), and the main phonon peak observed in Raman spectra (P2 mode, reproduced in Figure 4d). These experiments confirm the TO nature of the P2 phonon mode. Therefore, we establish that the P2 TO mode is the only phonon mode which couples to the electronic transition, whereas no contribution from LO phonons is found. The absence of the commonly observed electron-LO-phonon coupling points to the uniqueness of a ferroelectric semiconductor.

To support our interpretation of experimental results, we theoretically model the phonon dispersion using DFT (Figure 4e) and reveal the atomic motions that underlie the TO phonon mode (Figure 4f). By computing the inner product of eigenvectors and phonon momentum, we can estimate the TO and LO character of each phonon mode at different momentum vector *q* in the phonon dispersion. Along the Γ−A direction (perpendicular to the *bc* plane), the 2.830 THz phonon mode (3.130 THz from THz-TDs) is essentially purely TO near the zone center Γ, corroborating our THz-TDS results that indicate strong TO character of the P2 phonon mode. The same mode (2,830 THz along Γ−A in DFT) becomes highly longitudinal along the Γ−Y direction (parallel to *b* axis), and shifts to 3.128 THz (Figure S10 shows a clear track of the P2 mode near the Γ point).

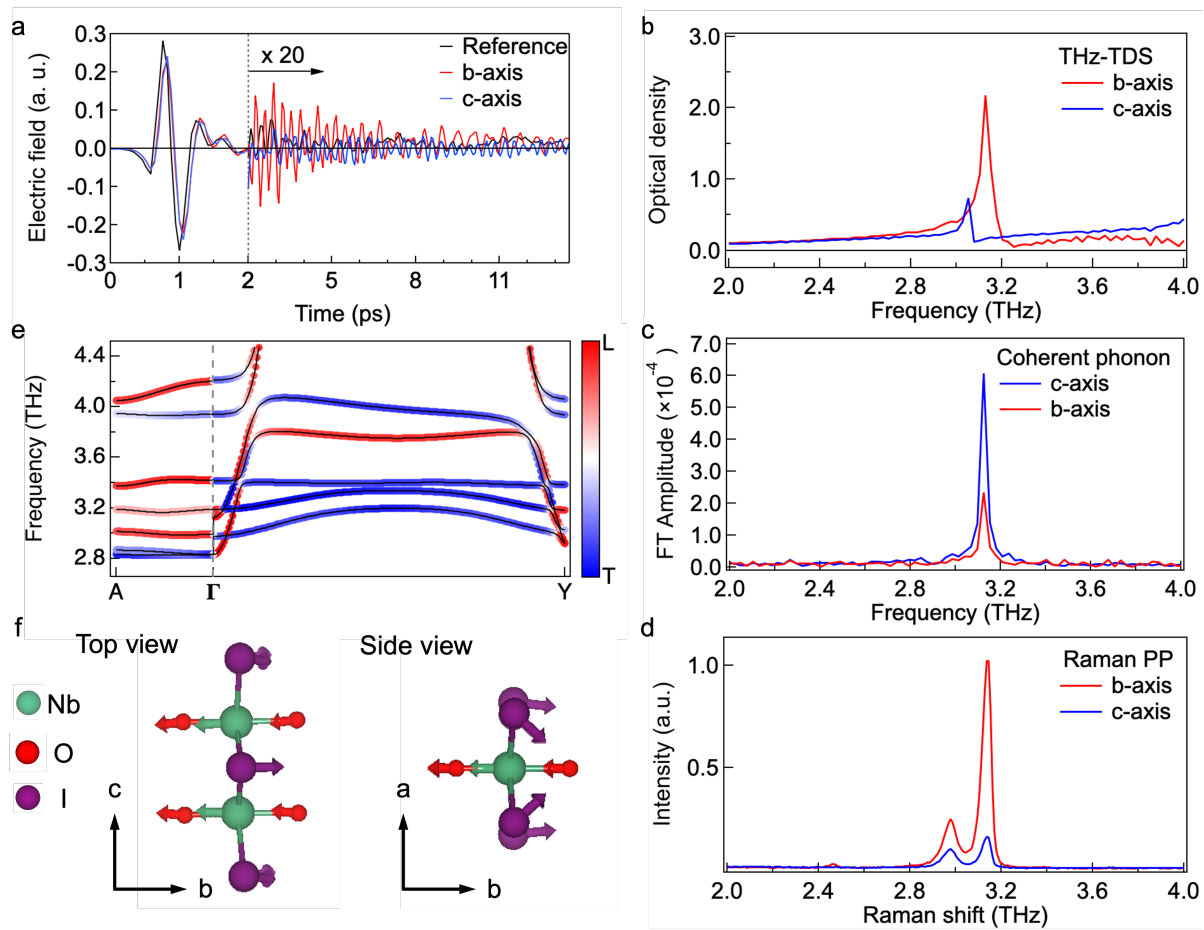

**Figure 4**. (**a,b**) Polarization-dependent THz absorption spectra of a 2.6-μm thick NbOI$_2$ in (**a**) time domain and (**b**) frequency domain, respectively. The reference signal represents the THz signal without transmitting through the NbOI$_2$ sample. The shoulder peak at 3.044 ± 0.023 THz does not match the shoulder peak found in the Raman spectrum with its Raman shift being 2.98 THz. We assign the 3.04 THz mode to a different TO phonon with relatively low Raman cross section which is nonetheless IR active. A full spectrum of THz absorption in frequency domain is shown in Fig. S9. **c.** FT spectra of the oscillatory component at exciton-resonant region obtained from the coherent phonon spectroscopy. **d.** Co-polarized Raman spectra of NbOI$_2$ with off-resonant pump. **e.** Calculated phonon dispersion based on first-principles DFT calculation with selected *q* directions. Here, **A** denotes a vector perpendicular to the crystallographic *bc*-plane, whereas **Y** denotes a vector parallel to crystallographic *b*-axis. Color contrast represents different levels of transverse or longitudinal character. **f.** Atomic motion of the TO phonon mode at 2.830 THz (calculated frequency) viewed along crystallographic *a*-axis (top view) and *c*-axis (side view).

The TO-LO splitting at zone center is therefore ~1.21 meV based on our calculations, manifesting the polar nature of this phonon mode along the direction associated with ferroelectric order.

Figure 4f shows the eigenvectors of the TO phonon mode. From both top and side views, we identify Nb and O atoms oscillating in-phase along the crystallographic *b*-axis, which modulates the local polarization. The calculated atomic motions are reminiscent of soft modes associated with displacive-type ferroelectrics; these soft modes are typically anharmonic along the direction associated with ferroelectric order. Although our calculations are based on the harmonic approximation, we expect anharmonicity in the TO phonon potential well due to its ferroelectric character. The asymmetric potential well as a result of this anharmonicity of the TO phonon will induce a nonvanishing change in the local polarization of the material when the phonon propagates. Such local polarization fluctuations can couple to the electronic transition and modulate the band energies near the lowest direct transition as manifested in our CPS results shown above.

To further support the distinct nature of the coupling of the P2 TO mode to the electronic transition, we performed Resonant Inelastic X-ray Scattering (RIXS) measurements[20–22] at the O-K edge (1s → 2p transition), Figure 5. RIXS probes electron-phonon coupling associated with the valence band and this coupling is expected to be distinctly different from that between electronic transition and the ferroelectric TO mode. Using linearly polarized incident photons parallel to the *b*-axis (π-polarization) and to the *c*-axis (σ-polarization), we find an anisotropic response in the XAS pre-edge (Figure 5b) and in the high energy loss region (> ~2eV) of the RIXS spectra (Figure 5a). This anisotropic behavior is consistent with the differential reflectance spectra in Figure 1c, and we associate it with the anisotropic electronic structure of $NbOI_2$. In particular, the anisotropic hybridization between O 2*p* and Nb 4*d* orbitals[32–35] is revealed by the strong anisotropic XAS feature around 530~535 eV[36]. In the low energy loss (< 50 meV) regime of the RIXS spectrum, several spectral components overlap in the vicinity of the elastic line and are visible with both σ and π polarization. When fitting the data (Figure 5c), we observe a resolution limited mode at 7 meV, together with its second-order harmonic at 14 meV, and its anti-Stokes counterpart at −7 meV. The 7 meV mode matches a phonon observed with THz spectroscopy (Figure S9) at 1.78 THz. The intensity ratio between the 7 and 14 meV modes and the evolution of their intensity as a function of incident photon energy supports our assignment of the second peak as the higher harmonic[33,35]. The presence of higher order phonon harmonics in the RIXS spectra signifies a non-negligible electron-phonon coupling between valence electrons and the 7 meV phonon mode[32–35]. While we cannot fully rule out a small contribution of the P2 phonon mode (~12.9 meV or 3.13 THz) in the RIXS response, the coupling of valence-band electron to phonon is dominated by the 7 meV phonon mode. Distinct from this valence-band electron-phonon coupling scenario, CPS results have revealed that the P2 phonon mode dominates coupling to above-gap electronic transition. Overall, the absence of a RIXS feature at ~3.13 THz supports our interpretation of the CPS data that points to an additional coupling mechanism beyond the common electron-phonon interaction associated with the valence band[37,38], in which a photo-excited carrier or above-gap optical transition is absent.

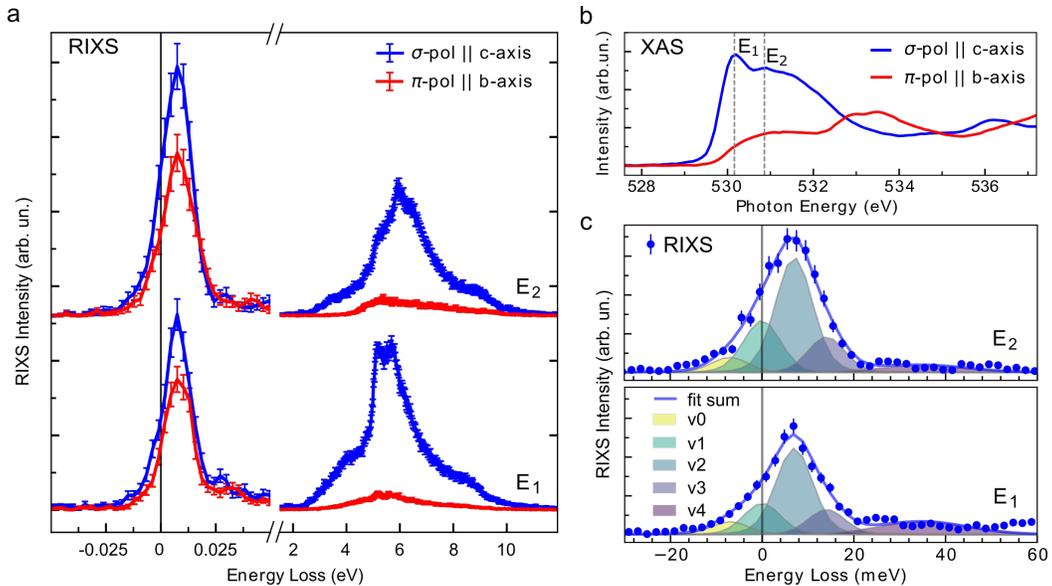

**Figure 5. a**. RIXS spectra measured at the O-K edge, with σ (blue) and π (red) polarized light, at T=35K. The incident photon energies were selected according to the main peaks displayed by the O-K XAS, shown in (**b**). **c**. Low-energy zoom-in of the RIXS spectra, measured for σ polarization (blue dots) at various incident photon energies. The spectra were fitted with five Voigt-shaped functions respectively centered at -7 meV (v0), 0 meV (v1), 7 meV (v2), 14meV (v3), and 36 meV (v4). The first four peaks are resolution limited. The fit sum is represented by the blue solid line. Similar results are obtained for π polarization, see Figure S11. RIXS spectra showing full range of energy loss were also shown in Figure S11.

**Conclusions**

We present experimental evidence for the dominant coupling of above-gap electronic transition to a phonon mode at 3.1 THz in the 2D ferroelectric semiconductor $NbOI_2$. This mode is identified as the TO phonon mode, uniquely linked to the ferroelectric polarization, by Raman spectroscopy, pump-probe coherent phonon spectroscopy, THz-TDS, and first principles calculation. Our findings provide direct evidence for the importance of specific forms of electron-TO phonon coupling in a 2D ferroelectric semiconductor. This is distinct from the well-known electron coupling to LO phonons in conventional semiconductors described by the Fröhlich model. The observed coupling to the TO phonon is specific to electronic excitation across the semiconductor band gap and is absent for valence band probed by RIXS spectroscopy. Our finding reveals the distinct coupling of above gap electronic transition and associated photo-carriers to ferroelectric polarization through the ferroelectric TO phonon mode. We suggest that this finding is of general importance to understanding electronic properties of 2D ferroelectric semiconductors where the interaction between TO phonon modes and electronic properties has been overlooked.

**Method:**

*Polarization-angle resolved Raman spectroscopy*

The low-frequency Raman scattering measurements were carried on a home-built setup with a Nikon TE-300 inverted microscopy. A 633-nm HeNe laser first passed a linear polarizer and was filtered by a reflective band-rejection filter (Ondax), then sent through a half wave plate to control the angle of the incident polarization, and into the microscope objective (40X, NA = 0.6). The incident light was focused on a cleaved $NbOI_2$ single crystal mounted on a $Si/SiO_2$ wafer secured in a vacuum environment (Oxford MicroStat Hi-Res2). The Raman scattered light was collected through the same objective and the same half wave plate, which projects the co-polarized Raman signal onto the polarization axis at the incident laser, and the cross polarized Raman signal onto the orthogonal axis. Then, the Raman scattered signal was sent through two notch filters to filter out the Rayleigh scattered laser line. The Raman signal was focused onto the entrance slit of a spectrometer (Princeton Instruments HRS-300) with a 2400 gr/mm holographic grating that dispersed the spectrum onto a $LN_2$ cooled CCD camera (Princeton Instruments LN400/B). The Raman shift was calibrated by a dual Hg/He atomic emission lamp.

*Polarization-angle resolved reflection spectroscopy*

Broadband tungsten-halogen lamp was used for light source. The broadband light was sent through a linear polarizer followed with a half wave plate to control the incident polarization. The polarized incident light was reflected to a 50:50 beam splitter with a small incident angle, and then sent through the same microscope used for polarization-angle resolved Raman spectroscopy and focused on an exfoliated few-layer $NbOI_2$ on a z-cut quartz substrate with the microscope objective (40X, NA = 0.6). The reflected white light was sent back through the same objective and beam splitter and was focused on the same spectrometer used for polarization-angle resolved Raman spectroscopy with a 150 gr/mm grating. The sample was measured under air.

*Coherent phonon spectroscopy*

The pump and probe beams were derived from the fundamental output of an ultrafast Ti:sapphire regenerative amplifier (Coherent Legend) operating at a 10 kHz repetition rate. The pump beam was converted to 480 nm (40 µW) by a home-built noncolinear optical parametric amplifier (NOPA) and mechanically chopped at 5 kHz to match the second resonant energy of $NbOI_2$ along the *c*-axis, while the probe beam was a broadband white light spanning 470-870 nm generated from a *c*-cut sapphire plate. The polarization of the incident pump and probe beams was controlled by half wave plates, with an additional linear polarizer used for the probe beam. The two beams overlapped at a nearly normal incidence on a $NbOI_2$ crystal (the same crystal used for polarization-angle resolved Raman spectroscopy) mounted on a $Si/SiO_2$ wafer in a nitrogen gas-tight cell. The reflected probe beam is then dispersed by a 600 gr/mm diffraction grating blazed at 500 nm and focused on a line-scan CCD camera (e2v AViiVA EM4). The beam spot diameter is ~ 200 µm.

*Nanosecond-scale transient reflectance spectroscopy*

Transient reflectance measurements were carried out by using the broadband pump-probe setup (Helios spectrometers, Ultrafast Systems). The pump beam of 470 nm was generated in a collinear optical parametric amplifier (OPerA Solo, Light conversion) pumped by the 800 nm output of an amplified Ti:Sapphire laser (Coherent Astrella, 1 kHz). The probe light was generated by focusing the fundamental of Astrella laser system in a sapphire crystal. The pump-probe delay was controlled by a mechanical delay line. The two beams overlapped at a nearly normal incidence on a NbOI$_2$ crystal (the same crystal used for polarization-angle resolved Raman spectroscopy and coherent phonon spectroscopy) mounted on a Si/SiO$_2$ wafer in a nitrogen gas-tight cell. The beam spot diameter is ~ 1 mm.

*Scanning tunneling microscopy (STM)*

STM measurements were performed on bulk crystals of NbOI$_2$ which were freshly cleaved in UHV. Measurements were conducted at T = 292 K with an Omicron VT-STM using electrochemically etched tungsten tips that were calibrated on the surface of Cu (111) prior to each sample approach. Topographic images were collected in the constant current mode with a setpoint bias and current of 1V, 50pA.

*Terahertz time-domain spectroscopy (THz-TDS)*

THz-TDS was performed based on the procedure described in ref. [39]. Briefly, the output of Ti:sapphire regenerative amplifier (RA) with a pulse duration of 30 fs, a repetition rate of 10 kHz, and a wavelength centered at 800 nm (Coherent, Legend) was separated into two beams for THz generation and sampling. The THz probe was generated using the two-color air plasma method. The THz beam was collimated and then focused onto an NbOI$_2$ sample mounted on a 1-mm precision pinhole (Thorlabs) that defines the sample area having high homogeneity. Reference TDS was taken with another blank precision pinhole. The transmitted THz probe was collected and focused onto a 1-mm (110) ZnTe crystal. Another 800-nm sampling beam was directed onto ZnTe, by which the time-domain THz field was recorded via electro-optic sampling. The transmittance was obtained by dividing the sample and reference data after Fourier transformation. Prior to THz-TDS, the crystal orientation was first determined using in-situ visible polarimetry equipped inside the THz setup. The THz setup was purged with dry air, and the measurements were performed at room temperature. For the NbOI$_2$ sample preparation for THz-TDS, a flake of NbOI$_2$ is cleaved form a bulk crystal with thermal release tape (TRT) and thinned with Scotch tape. The NbIO$_2$ thin flake is mounted to a pinhole with double sided Kapton tape and heated to 135C to remove the TRT.

*First-principles calculation*

The phonon and Raman spectra were simulated using first-principles density functional theory (DFT) calculations. We employed the Perdew-Burke-Ernzerhof (PBE) functional[40] for electronic structure and atomic position optimization, as implemented in the VASP package[41,42]. The simulation was conducted using the primitive cell of $NbOI_2$, with a plane-wave basis cutoff of 500 eV. The Brillouin zone was sampled using a Γ-centered 6×6×6 k-point mesh. The force on each atom was converged to less than 0.0001 eV/Å. Phonon calculations were performed using the supercell frozen phonon approach, with structures generated by the Phonopy code[43,44] based on a 3×3×3 supercell. Non-analytic corrections for LO-TO splitting in polar materials were accounted for using Born effective charges. The dielectric tensor and Born effective charges were calculated through density functional perturbation theory (DFPT)[45]. Raman tensors were subsequently derived using finite differences on the dielectric tensors[46]. For experimental comparison, phonon linewidths were computed via the 3rd order force constants [43,47] and a supercell approach, involving 9240 displaced structures from a 2×2×2 supercell. The final angular-dependent Raman intensities included phonon linewidth at the Γ point at 300 K.

*X-ray Absorption Spectroscopy (XAS) and Resonant Inelastic X-ray Scattering (RIXS)*

The XAS and RIXS measurements were performed at the SIX 2-ID Beamline of NSLS-II[48]. The RIXS energy resolution was $\Delta E = 13$ meV. The spectrometer was set to $2\theta = 150°$, while the sample incident angle was $\theta = 90°$. Under this experimental configuration, the σ-polarized incoming x-rays were parallel to the sample *c*-axis, while the π-polarized ones were parallel to the *b*-axis. The temperature was 35K for all measurements. The $NbOI_2$ single crystal, aligned using the reflectance measurement, was glued by silver paint on a Cu-sample plate and cleaved in air with scotch-tape just before loading it into the loadlock chamber.

**Acknowledgment:**

The Raman and coherent phonon spectroscopy experiments are supported by the US Army Research Office, grant number W911NF-23-1-0056. The THz-TDS and DFT efforts are supported by the Materials Science and Engineering Research Center (MRSEC) through NSF grant DMR-2011738. Support by the Air Force Office of Scientific Research under award number FA9550-22-1-0389 is acknowledged for the imaging experiments. The RIXS experiments were supported by the Center on Programmable Quantum Materials, an Energy Frontier Research Center funded by the U.S. Department of Energy (DOE), under award number DE-SC0019443. This research used beamline 2-ID of NSLS-II, a US DOE Office of Science User Facility operated for the DOE Office of Science by Brookhaven National Laboratory under contract no. DE-SC0012704. C.-Y. H. is supported by the Taiwan-Columbia Fellowship funded by the Ministry of Education of Taiwan and Columbia University. M.Y.S was supported by the Gordon and Betty Moore Foundation, grant DOI 10.37807/gbmf12235. We thank Prof. Dr. Sebastian Maehrlein for fruitful discussions.

**Author contributions:**

C.-Y.H., X.R., and X.-Y.Z. conceived this project. C.-Y.H. carried out optical measurements with assistance by T.H., Y.L., G.H, and M.S. C.-Y.H. and N.O. prepared the exfoliated samples. D.G.C. synthesized the crystals and performed PXRD under the supervision by X.R. T.H. carried out THz-TDS measurements. D.G.C. and C.-Y.H built the high-temperature optical measurement setup. Z.-H.C. carried out theoretical calculation under the supervision by D.R.R. M.T. performed the STM under the supervision by A.P. Y.S performed NIR reflectance measurements under the supervision by D.N.B. M.E.Z. proposed the material. C.A.O., J.P, and V.B. carried out the XAS and RIXS. The manuscript was prepared by C.-Y.H., D.G.C., Z.-H.C., V.B., D.R.R., X.R., and X.-Y.Z. in consultation with all other authors. All authors read and commented on the manuscript.

# Supporting Information

# Coupling of Electronic Transition to Ferroelectric Order in a 2D Semiconductor

Chun-Ying Huang,[1,*] Daniel G. Chica,[1,*] Zhi-Hao Cui,[1,*] Taketo Handa,[1] Morgan Thinel,[1,2] Nicholas Olsen,[1] Yufeng Liu,[1] Michael E. Ziebel,[1] Guiying He,[3,4] Yinming Shao,[2,6] Connor A. Occhialini,[2,5] Jonathan Pelliciari,[5] Dmitri N. Basov,[2] Matthew Sfeir,[3,4] Abhay Pasupathy,[2] Valentina Bisogni,[5] David R. Reichman,[1,†] Xavier Roy,[1,†] and Xiaoyang Zhu.[1,†]

[1]Department of Chemistry, Columbia University, New York, NY 10027, USA
[2]Department of Physics, Columbia University, New York, NY 10027, USA
[3]Department of Physics, Graduate Center, City University of New York, New York, NY 10016, USA
[4]Photonics Initiative, Advanced Science Research Center, City University of New York, New York, NY 10031, USA
[5]National Synchrotron Light Source II, Brookhaven National Laboratory, Upton, NY 11973, USA
[6]Department of Physics, Pennsylvania State University, University Park, PA, 16802, USA

This PDF file includes:

Supplementary Text. *Raman Tensor Analysis.*
Supplementary Figures S1-S12
Supplementary Table S1
Supplementary Method
References

[*] These authors contributed equally.
[†] To whom correspondences should be addressed. DRR: drr2103@columbia.edu; XR: xr2114@columbia.edu; XYZ: xyzhu@columbia.edu

**Supplementary Text**

*Raman Tensor Analysis.* Since $NbOI_2$ has $C_2$ space group, we write down the Raman intensity of phonon modes with $A$ symmetry as a function of the angle of incident polarization in equations (1) and (2) for parallel and cross polarization, respectively[1-4]. The corresponding equations for $B$ symmetry under parallel and cross polarization are shown in equations (3) and (4), respectively[1-4].

$$I_{A,\|} \propto |c|^2\cos^4\theta + |b|^2\sin^4\theta + 2|b||c|\cos^2\theta\sin^2\theta\cos\varphi_{c-b} + bg \quad (1)$$

$$I_{A,\perp} \propto \frac{1}{4}(|c|^2 + |b|^2 - 2|b||c|\cos\varphi_{c-b})\sin^2 2\theta + bg \quad (2)$$

$$I_{B,\|} \propto |f|^2\sin^2 2\theta + bg \quad (3)$$

$$I_{B,\perp} \propto |f|^2\cos^2 2\theta + bg \quad (4)$$

The polarization dependences of the P2 phonon mode (Figure 2d) is well described by fits (solid curves) to equations (1) and (2) under $A$ symmetry, not $B$ symmetry (see Figure S5). We included a (isotropic) constant in the fitting. The tensor fittings of the remaining phonon modes are shown in Figure S5 where they are either $A$ or $B$ symmetry, as suggested by $C_2$ space group.

**Supplementary Figures**

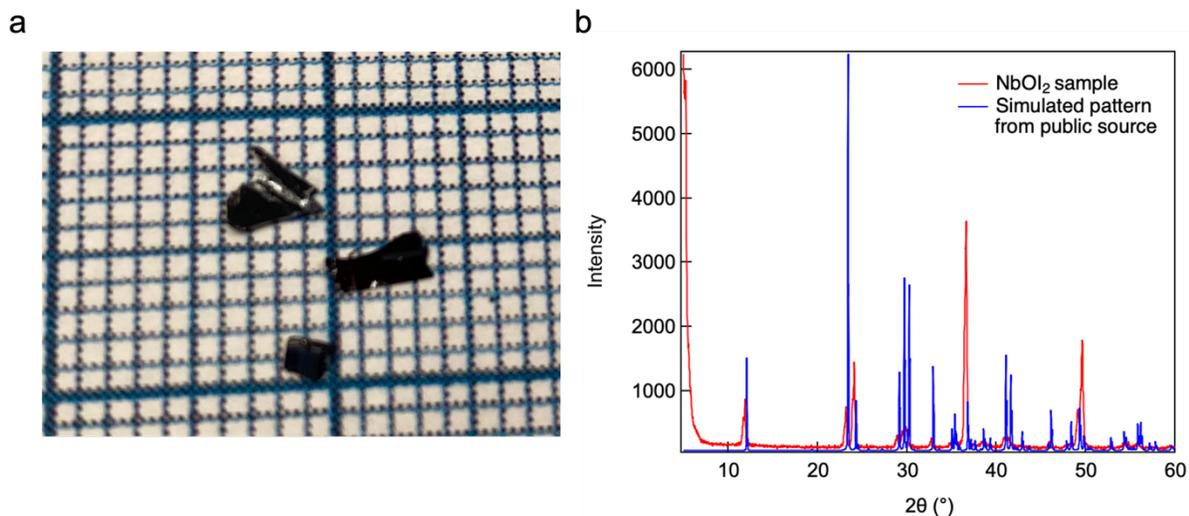

**Figure S1 a.** Photo of the grown $NbOI_2$ placed on millimeter grid sheet. **b.** Powder X-ray diffraction (PXRD) of as grown $NbOI_2$ single crystal. Simulated spectrum was generated by the crystal structure reported in ref. [5].

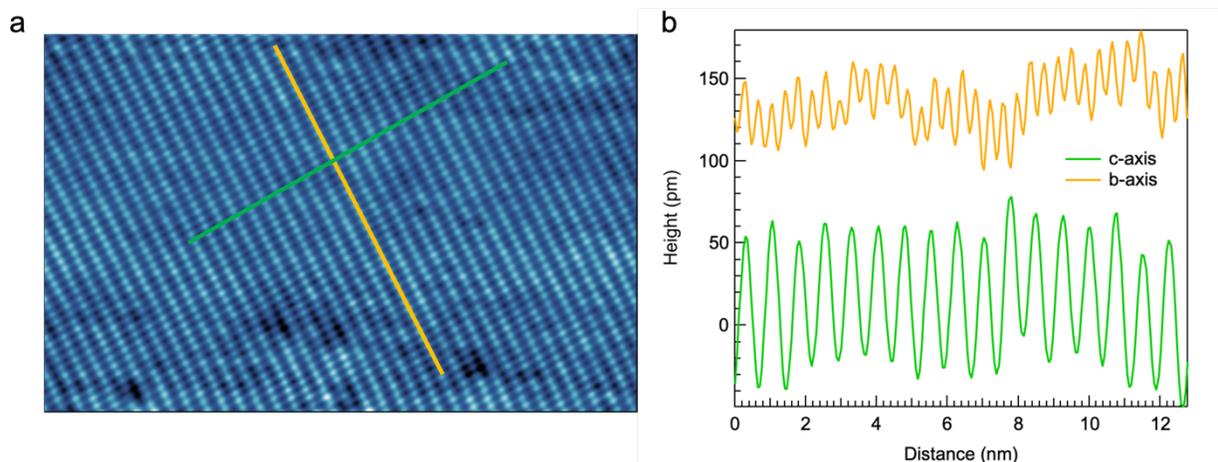

**Figure S2 a.** Scanning tunneling microscopy image of the cleaved NbOI$_2$ surface labeled with line traces corresponding to the *c*- and *b*-axis. **b.** Height profiles of the line traces shown in the left image give lattice constants of 0.388 and 0.741 nm for the *b*- and *c*-axis, respectively. The lattice constants are consistent with those reported in ref. 5.

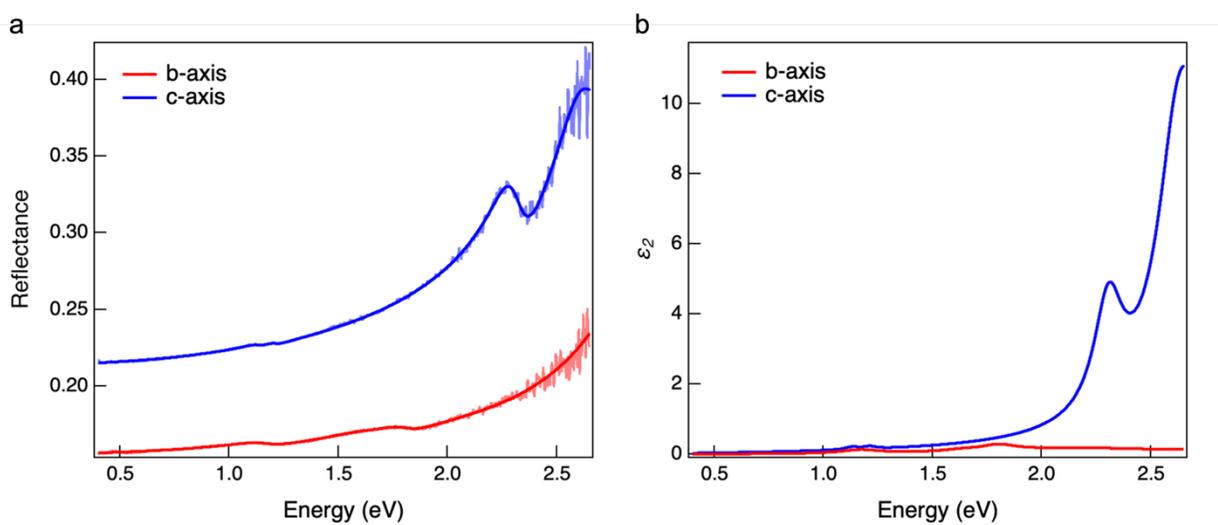

**Figure S3 a.** Extended reflectance spectra of bulk NbOI$_2$ down to Infrared range with incident light polarized along either the *c*- or *b*-axis at RT. Lighter lines are raw data points and darker lines are fitting curves and are for guidance of reading. **b.** Extracted imaginary dielectric constants of NbOI$_2$. Onset of an indirect-gap can be found around ~0.5 eV. The indirect-gap feature and spontaneous polarization of NbOI$_2$ suggest the formation of photo-excited carriers instead of exciton[6].

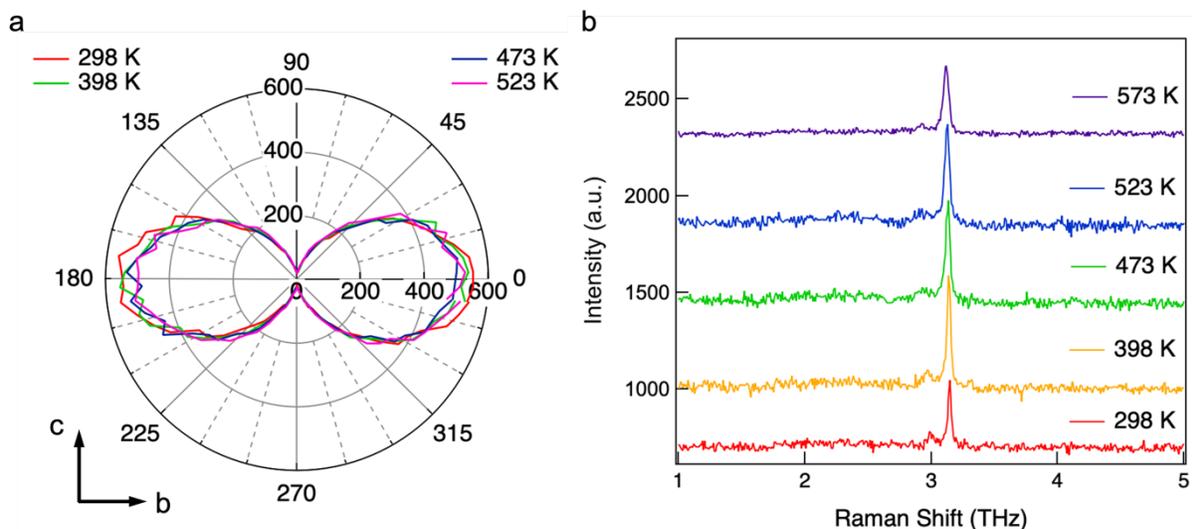

**Figure S4 a.** Temperature-dependent angle-resolved SHG intensity of bulk $NbOI_2$ as a function of the rotation angle of light polarization with respect to crystallographic *b*-axis. **b.** Temperature-dependent Raman spectra of bulk $NbOI_2$. Incident light polarization angle is selected to optimize the Raman intensity of 3.137 THz mode (P2 mode). The $NbOI_2$ crystal are sealed in a quartz tube filled with 0.5 atm of Ar gas at 298K to avoid off gassing and decomposition. Our result differs from the reported phase transition temperature in ref 7, which reports the ferroelectric-to-paraelectric phase transition to be ~463 K[7].

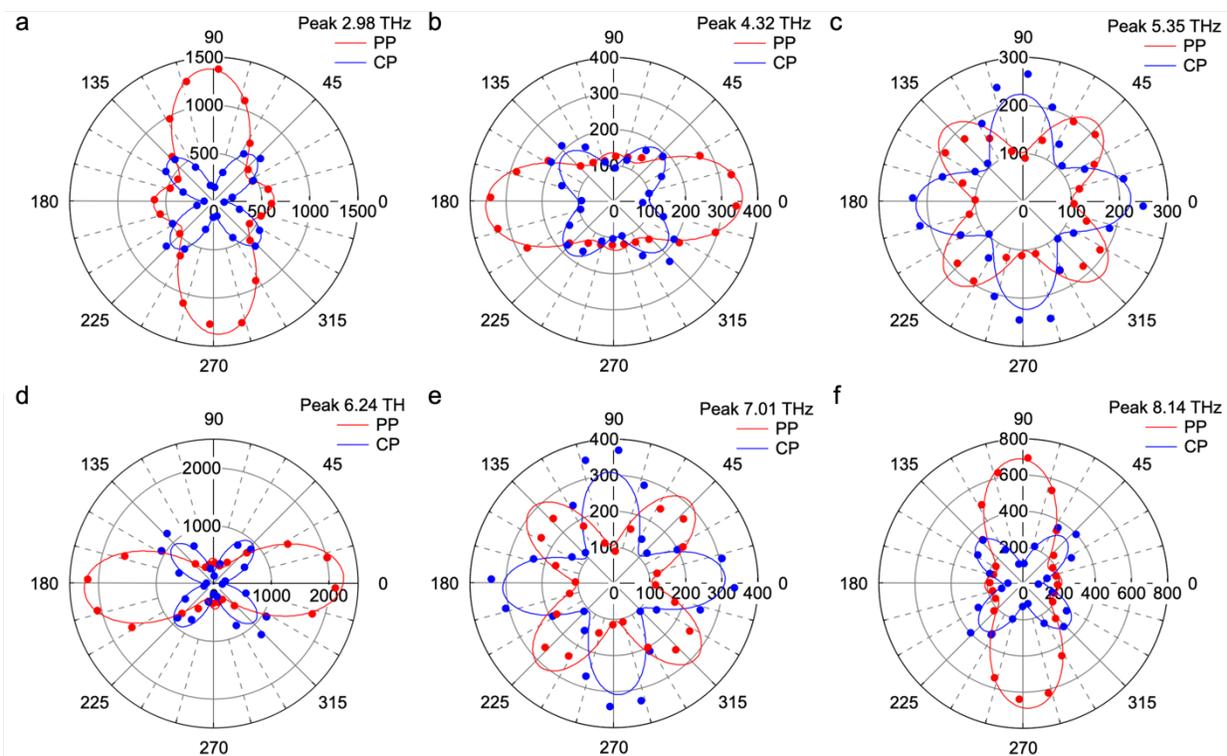

**Figure S5** Full set of tensor analysis for the observed modes. **(a, b, d, f)** Phonon modes that can be described by *A*-symmetry. **(c, e)** Phonon modes that can be described by *B*-symmetry. Phonon

modes under *B*-symmetry show intensity deviation from fitting curves. This is potentially due to the relative low intensity and signal noise ratio.

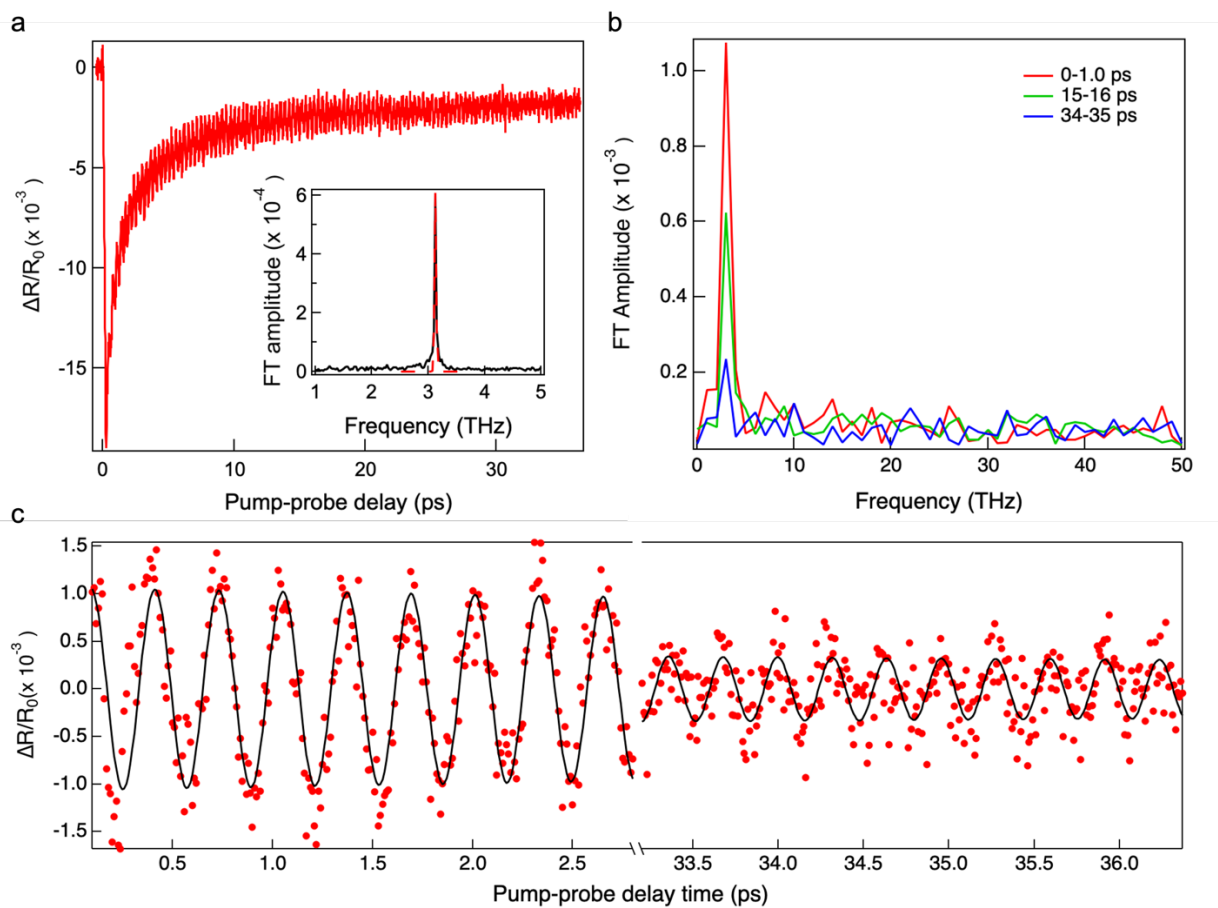

**Figure S6 a.** Transient reflectance probed at 2.30 eV as a function of pump-probe delay. Inset: Fourier transform (FT) of the oscillatory component (black line) and the red dash line is a Gaussian fit giving a peak position of $3.125 \pm 0.001$ THz and a FWHM of $0.038 \pm 0.001$ THz. **b.** Short-time Fourier transform (STFT) of the oscillatory component probed at 2.30 eV. A window size of 100 (1 ps) is used. The frequency of the 3.1 THz remains the same throughout the whole probing window. **c.** Fit (black line) of dephasing time of the oscillation amplitude probed at 2.30 eV gives an exponential decay of $29.32 \pm 2.08$ ps.

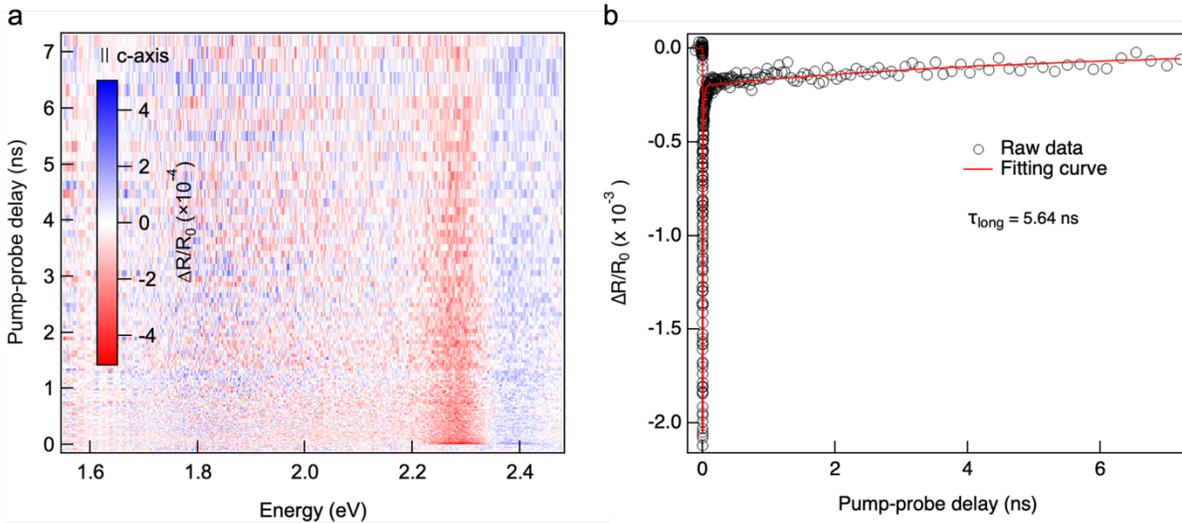

**Figure S7 a.** Transient reflectivity spectrum of NbOI$_2$ at RT with delay time extended to ~ 7 nanosecond (ns). The polarization of probe light is parallel to the *c*-axis. **b.** Kinetic trace of exciton bleaching extracted from the left color map. The intensity is integrated within 2.24 – 2.33 eV region. We use multiexponential reconvolution fit with a gaussian function to extract the long-decay component $\tau_{long}$ = 5.6 ± 1.2 ns.

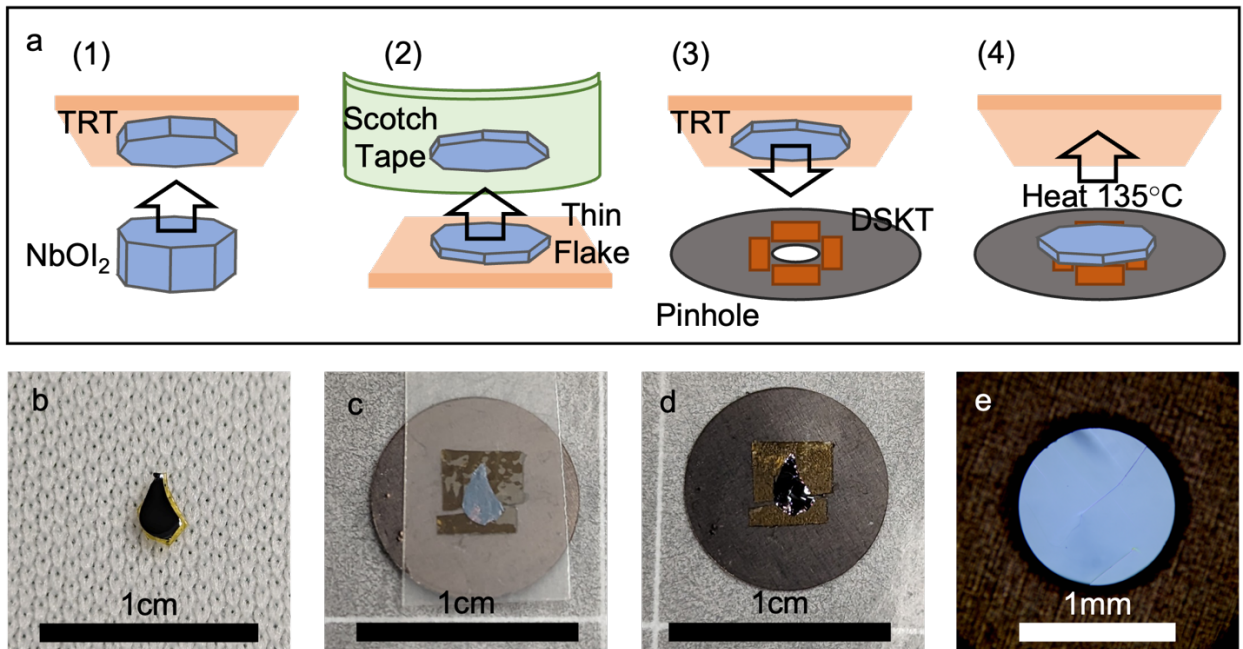

**Figure S8 a.** Fabrication schematic for the preparation of thin flakes of NbOI$_2$ suspended over a pinhole. (1) Bulk NbOI$_2$ crystal is cleaved with thermal release tape (TRT); (2) the NbOI$_2$ flake on the TRT is thinned by cleaved with Scotch tape; (3) the thinned NbOI$_2$ flake is mounted to a pinhole with double sided Kapton tape (DSKT); (4) the sample assembly is heated to 135°C to remove the thermal release tape. **b.** Image of bulk NbOI$_2$ crystal mounted to a glass microscope slide with DSKT. Excess DSKT has been removed such that only the bulk crystal is exposed. **c.**

Image of a NbOI$_2$ thin flake mounted on a pinhole with DSKT before TRT removal. **d.** Image of a NbOI$_2$ thin flaked mounted on the pinhole after TRT removal. **e.** Optical microscope image of the exposed area of the NbOI$_2$ thin flake taken through the pinhole. The crystal thickness is determined based on the method reported in ref. (*Taketo's manuscript submitted to Nat Mat*)

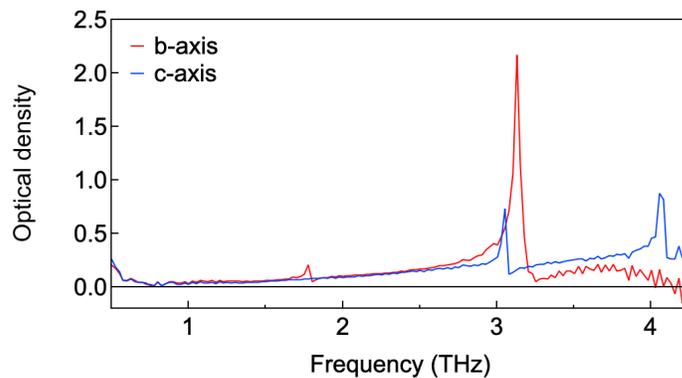

**Figure S9.** Full spectrum of polarization-dependent THz absorption in frequency domain.

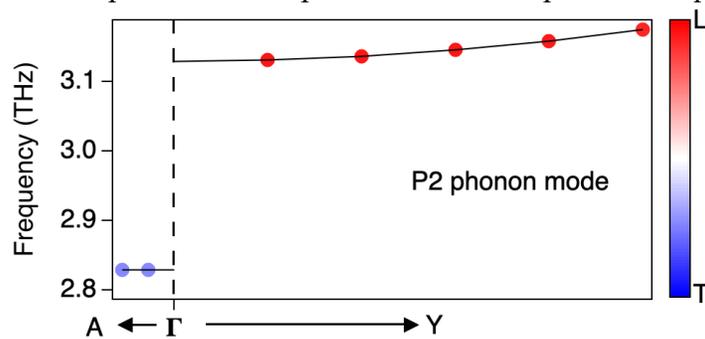

**Figure S10.** Phonon dispersion of the P2 phonon mode near the Γ point along the A- Γ-Y direction.

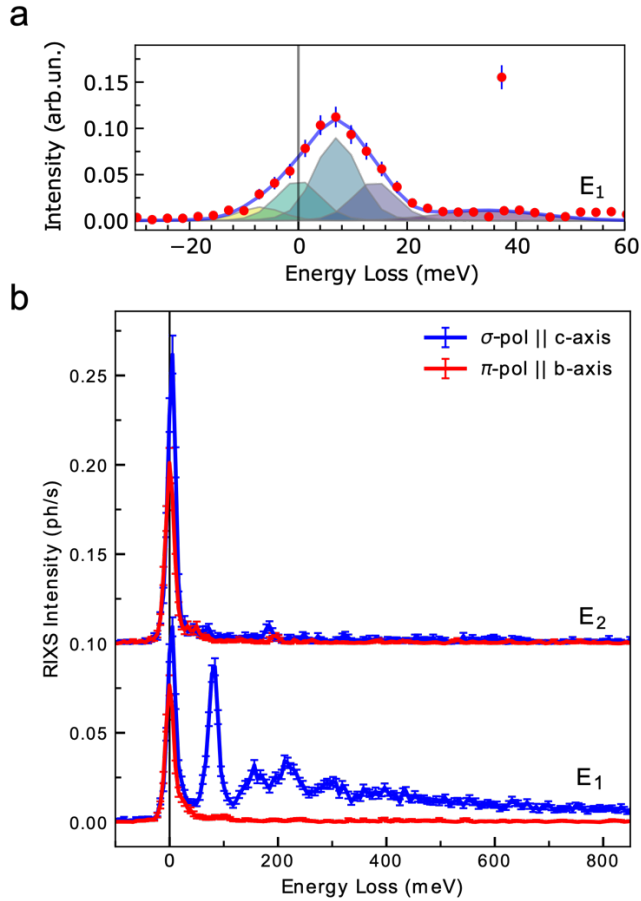

**Figure S11. a.** Low-energy zoom-in of the RIXS spectra, measured for π polarization (red dots) at various incident photon energies. The spectra were fitted with five Voigt-shaped functions respectively centered at -7 meV (v0), 0 meV (v1), 7 meV (v2), 14meV (v3), and 36 meV (v4). The first four peaks are resolution limited. The fit sum is represented by the blue solid line. **b.** RIXS spectra with Full range of energy loss. The phonon mode at 80 meV is detectable only when the resonant energy is at first XAS peak ($E_1$) with σ polarized light. This suggests that the 80 meV phonon mode is selectively coupled to a specific state hybridizing O *2p* and Nb *4d* orbitals.

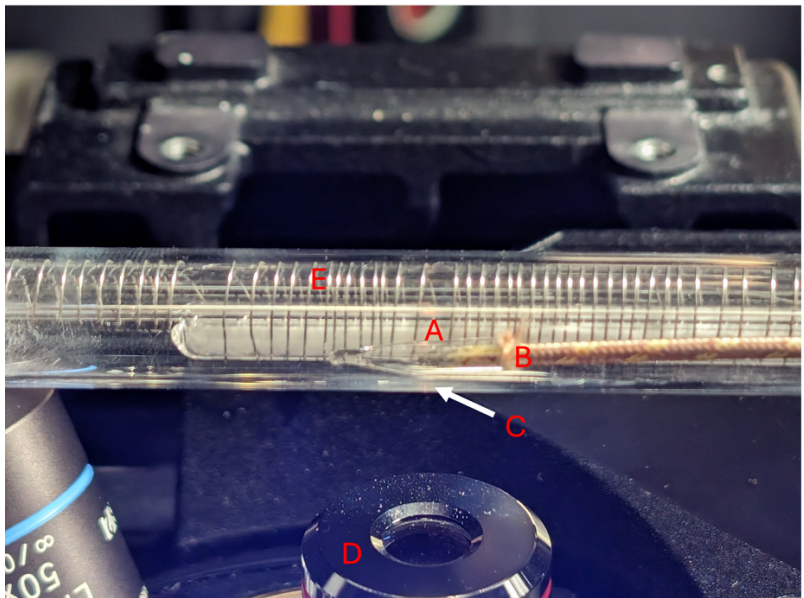

**Figure S12.** Picture of the setup for the high-temperature SHG/Raman measurement. A: quartz ampoule with an $NbOI_2$ crystal. B: K-type thermocouple. C: 1-mm aperture. D: 10X microscope objective. E: nichrome heating element.

**Supplementary Table**

**Table S1.** Calculated Raman-active phonon modes of NbOI$_2$ with their symmetry assignment at RT.

| Phonon mode | Calculated frequency (THz) | Experimental frequency (THz) | Raman tensor* | Symmetry |
|---|---|---|---|---|
| 1 | 0 | | $\begin{bmatrix} 0.00000000 & 0.00000000 & 0.00027061 \\ 0.00000000 & -0.00036081 & 1 \\ -0.00027061 & 0.00000000 & 0.00027061 \end{bmatrix}$ | A |
| 2 | 0 | | $\begin{bmatrix} 0.00378853 & 0.00000000 & -0.00487096 \\ 0.00000000 & -0.00027061 & 0.00000000 \\ -0.00460035 & 0.00000000 & 0.00496117 \end{bmatrix}$ | A |
| 3 | 0 | | $\begin{bmatrix} 0.00423954 & 0.00000000 & -0.00514157 \\ 0.00000000 & -0.00090203 & 0.00000000 \\ -0.00541218 & 0.00000000 & 0.00667502 \end{bmatrix}$ | A |
| 4 | 1.614 | | $\begin{bmatrix} -0.00007626 & -0.25655262 & 0.00091517 \\ -0.25670514 & 0.00030506 & -0.18135616 \\ 0.00076264 & -0.18143242 & -0.00114396 \end{bmatrix}$ | B |
| 5 | 1.788 | | $\begin{bmatrix} 0.04679394 & 0.0000000 & 0.00129983 \\ 0.00000000 & 0.60472762 & 0.00000000 \\ 0.00145275 & 0.00000000 & 0.03868911 \end{bmatrix}$ | A |
| 6 | 1.960 | | $\begin{bmatrix} 0.00199373 & -1.16180725 & 0.00437087 \\ -1.16173056 & -0.00007668 & -0.28932075 \\ 0.00582782 & -0.28932075 & 0.00092018 \end{bmatrix}$ | B |
| 7 | 2.381 | | $\begin{bmatrix} -0.74057732 & 0.00000000 & 0.26290136 \\ 0.00000000 & 0.17347180 & 0.00000000 \\ 0.26317072 & 0.00000000 & 0.45343301 \end{bmatrix}$ | A |
| 8 | 2.829 | 3.137 | $\begin{bmatrix} -1.69304490 & 0.00000000 & 0.26142054 \\ 0.00000000 & -6.90720314 & 0.00000000 \\ 0.26010781 & 0.00000000 & -3.23250056 \end{bmatrix}$ | A |
| 9 | 2.832 | | $\begin{bmatrix} -0.00177282 & 0.09854806 & -0.00062570 \\ 0.09886091 & -0.00125140 & -0.56188038 \\ -0.00177282 & -0.56459176 & -0.00594417 \end{bmatrix}$ | B |
| 10 | 2.990 | 2.978 | $\begin{bmatrix} 0.50331098 & 0.00000000 & 0.59698102 \\ 0.00000000 & -1.64977587 & 0.00000000 \\ 0.59671617 & 0.00000000 & 0.20102422 \end{bmatrix}$ | A |
| 11 | 3.188 | | $\begin{bmatrix} 0.36672827 & 0.00000000 & 0.09816381 \\ 0.00000000 & 0.07973772 & 0.00000000 \\ 0.09840841 & 0.00000000 & -0.23285370 \end{bmatrix}$ | A |
| 12 | 3.415 | | $\begin{bmatrix} -0.00226615 & 0.02602167 & -0.00453230 \\ 0.02649053 & 0.00023443 & -0.11315129 \\ -0.00672031 & -0.11362014 & 0.00093772 \end{bmatrix}$ | B |
| 13 | 3.941 | | $\begin{bmatrix} 0.00354701 & -0.17790212 & 0.00189174 \\ -0.17837506 & 0.00102469 & 0.53819924 \\ 0.00055176 & 0.53654397 & 0.00031529 \end{bmatrix}$ | B |
| 14 | 4.202 | 4.334 | $\begin{bmatrix} -5.69638720 & 0.00000000 & -1.55040470 \\ 0.00000000 & -0.33442634 & 0.00000000 \\ -1.55161551 & 0.00000000 & -1.42988852 \end{bmatrix}$ | A |
| 15 | 4.941 | 5.331 | $\begin{bmatrix} 0.00357781 & 0.62192841 & -0.00078537 \\ 0.62349915 & 0.00000000 & -2.10549861 \\ -0.00017453 & -2.09755762 & -0.00191980 \end{bmatrix}$ | B |
| 16 | 5.951 | | $\begin{bmatrix} 2.80243138 & 0.00000000 & -3.89992384 \\ 0.00000000 & 0.19961587 & 0.00000000 \\ -3.89825672 & 0.00000000 & -2.56420848 \end{bmatrix}$ | A |
| 17 | 6.225 | 6.236 | $\begin{bmatrix} 2.27562944 & 0.00000000 & -1.84736169 \\ 0.00000000 & -0.90160930 & 0.00000000 \\ -1.84789354 & 0.00000000 & 6.06435633 \end{bmatrix}$ | A |
| 18 | 6.630 | 7.005 | $\begin{bmatrix} -0.01051776 & 1.05139393 & -0.00726682 \\ 1.05521857 & 0.00057370 & -2.70555135 \\ -0.01013530 & -2.69618098 & -0.00382464 \end{bmatrix}$ | B |
| 19 | 6.646 | | $\begin{bmatrix} -0.00737552 & 0.23707024 & -0.01053646 \\ 0.23865071 & 0.00000000 & -2.29448870 \\ -0.01492664 & -2.28693757 & -0.00456580 \end{bmatrix}$ | B |
| 20 | 6.685 | | $\begin{bmatrix} 0.54602812 & 0.00000000 & -1.33595989 \\ 0.00000000 & -0.00436656 & 0.00000000 \\ -1.33866300 & 0.00000000 & -0.39236674 \end{bmatrix}$ | A |
| 21 | 8.000 | 8.142 | $\begin{bmatrix} -0.06966349 & 0.00000000 & -0.24953535 \\ 0.00000000 & -1.74214008 & 0.00000000 \\ -0.24953535 & 0.00000000 & -1.20252449 \end{bmatrix}$ | A |
| 22 | 8.113 | | $\begin{bmatrix} -0.00716349 & -1.44069485 & -0.01349402 \\ -1.44219419 & 0.00049978 & -0.41015156 \\ -0.01132831 & -0.41065134 & -0.00183252 \end{bmatrix}$ | B |
| 23 | 14.218 | | $\begin{bmatrix} -0.00352986 & -0.87125183 & 0.00394513 \\ -0.87166711 & 0.00083055 & 3.02051965 \\ 0.00809791 & 3.01948145 & -0.02720066 \end{bmatrix}$ | B |
| 24 | 14.335 | | $\begin{bmatrix} 6.31891783 & 0.00000000 & -5.35411870 \\ 0.00000000 & 48.11290064 & 0.00000000 \\ -5.35184615 & 0.00000000 & 24.91041118 \end{bmatrix}$ | A |

*The coordinate of Raman tensors is based on primitive cell of $NbOI_2$ instead of unit cell. The high-symmetry C$2$ axis is aligned with the *b*-axis of $NbOI_2$.

**Supplementary method**

*Crystal growth of NbOI₂*

Niobium powder (0.0768 grams, Thermo Fischer Scientific, Puratronic 99.99% purity), niobium(v) oxide powder (0.0733 grams, Thermo Fischer Scientific, Puratronic 99.9985% purity), and iodine chunks (0.3649 grams, Sigma Aldrich, 99.99+% purity) were loaded into a 12.7mm o.d., 10.5 mm i.d. fused silica tube backfilled with argon. This ratio of reagents corresponds to 0.5000 grams of $NbOI_2$ with an additional 0.0150 grams of $I_2$ to facilitate crystal growth via a chemical vapor transport reaction. The tube was evacuated to a pressure of ~30 mtorr and submerged under liquid $N_2$ to prevent the volatilization of iodine. The tube was sealed to a length of 12 cm. Two heating methods were used which both effectively produced the same $NbOI_2$ material. In the first heating method, the ampoule was placed horizontally into a computer-controlled box furnace with the precursor material towards the back of the furnace and the other side of the tube towards the furnace door. The tube was heated with the following heating program: Heat to 220°C in 20 minutes, dwell for 24 hours, heat to 600°C in 48 hours, dwell for 72 hours, and then turn off furnace and cool to ambient temperature radiatively. Crystals of $NbOI_2$ grew on the tube end closest to the door from the natural temperature gradient that occur inside box furnaces. For the second heating method, a more deliberate temperature gradient was applied during the heat treatment through use of a computer controlled two-zone furnace. The ampoule was prepared the same way as above except the following mass of reagents were loaded: Nb (0.1537 grams), $Nb_2O_5$ (0.1466 grams), and $I_2$ (0.7147 grams). These masses correspond to 1.0000 grams of $NbOI_2$ and 0.0150 grams of $I_2$. The tube was heated in computer controlled two zone furnaces with the following heating profile where the source and sink side are the locations of the reagents and deposited $NbOI_2$ crystals, respectively. Source side heating profile: Heat to 220°C in 3 hours, dwell for 24 hours, heat to 500°C in 24 hours, dwell for 24 hours, heat to 550°C in 6 hours, dwell for 144 hours, cool to ambient temperature in 6 hours. Sink side heating profile: Heat to 220°C in 3 hours, dwell for 24 hours, heat to 550°C in 24 hours, dwell for 24 hours, heat to 500°C in 6 hours, dwell for 144 hours, cool to ambient temperature in 6 hours. The excess iodine on the crystal surface was removed by heating the sink side with the crystal to ~100°C and keeping the source at ambient temperature.

*Powder X-ray diffraction*

Crystals of $NbOI_2$ were lightly ground and then mounted on a zero-background silicon holder. Powder diffraction patterns were collected using a Marvin Panalytical diffractometer with a Cu Kα X-ray source powered to 40 kV and 15 mA and filtered by a Niβ filter. The sample was spun while collecting to reduce the effects of preferred orientation. The collected $NbOI_2$ diffraction pattern was then compared with a simulated diffraction pattern generated from the crystal structure reported from ref. 1.

*High temperature optical measurements*

A crystal of $NbOI_2$ was loaded into a 3x3mm square fused silica tube (0.5 mm thickness) and affixed with quartz wool such that the normal to the vdW plane was parallel to the planar quartz wall. The tube was sealed under 0.5 atm of argon gas. The sealed ampoule was assembled into a custom heater set up as seen in Fig S12. Nichrome wire was wrapped around the ampoule and a K-type thermocouple was placed next to the wire/ampoule assembly which were both surrounded by an outer fused quart tubing. The outer tubing had a ~ 1mm aperture to prevent scattering of the SHG pump and signal. The thermocouple and heating element were connected to a Digi-Sense TC6000 thermocouple temperature control box to stabilize temperatures ranging from ambient temperature up to 300°C. The same optical set up to collect room temperature SHG data was used with this heating apparatus to collect high temperature SHG data.

*Second harmonic generation (SHG) measurements*

SHG is carried out with Spectra-Physics Tsunami 80 MHz Ti:Sapphire oscillator that generates 800 nm to excite the sample. The incident beam is first reflected by a 650-nm short-pass dichroic mirror, sent through an achromatic half waveplate, and then focused onto the sample with a 10x objective. A polarizer is inserted before the dichroic mirror to ensure the polarization purity of the incident light. The SHG signal transmits the dichroic mirror and enters photomultiplier tube. An additional short pass filter is used to reduce residual fundamental light.

**Supplementary References**